\documentclass[%
 aip,
 amsmath,amssymb,
 reprint,%
]{revtex4-1}

\usepackage{graphicx}
\usepackage{dcolumn}
\usepackage{bm}

\usepackage[utf8]{inputenc}
\usepackage[T1]{fontenc}
\usepackage{etoolbox}

\makeatletter
\def\@email#1#2{%
 \endgroup
 \patchcmd{\titleblock@produce}
  {\frontmatter@RRAPformat}
  {\frontmatter@RRAPformat{\produce@RRAP{*#1\href{mailto:#2}{#2}}}\frontmatter@RRAPformat}
  {}{}
}%
\makeatother
\begin{document}

\preprint{AIP/123-QED}

\title{Brownian motion of a rod threading through a ring with fixed ring-center}
\author{Zhongqiang Xiong}
\affiliation{ 
Zhejiang Key Laboratory of Soft Matter Biomedical Materials, Wenzhou Institute, University of Chinese Academy of Sciences, Wenzhou, Zhejiang 325000, China
}
\affiliation{ 
Oujiang Laboratory (Zhejiang Lab for Regenerative Medicine, Vision and Brain Health), Wenzhou 325000, China
}
\author{Shigeyuki Komura}%
\affiliation{ 
Zhejiang Key Laboratory of Soft Matter Biomedical Materials, Wenzhou Institute, University of Chinese Academy of Sciences, Wenzhou, Zhejiang 325000, China
}
\affiliation{ 
Oujiang Laboratory (Zhejiang Lab for Regenerative Medicine, Vision and Brain Health), Wenzhou 325000, China
}

\author{Masao Doi}
\email{doi.masao.y3@a.mail.nagoya-u.ac.jp}
\affiliation{ 
Zhejiang Key Laboratory of Soft Matter Biomedical Materials, Wenzhou Institute, University of Chinese Academy of Sciences, Wenzhou, Zhejiang 325000, China
}
\affiliation{ 
Oujiang Laboratory (Zhejiang Lab for Regenerative Medicine, Vision and Brain Health), Wenzhou 325000, China
}

\date{\today}

\begin{abstract}

We study the Brownian motion of a rigid rod threading through a small fixed ring while the ring can freely rotate. We derive the distribution function for the sliding displacement and the unit vector along the rod both at equilibrium and non-equilibrium. The equilibrium distribution is quadratic in the sliding displacement and is controlled by the moment of inertia (mass distribution). Applying the Onsager variational principle, we derive a Smoluchowski equation in which sliding and rotational diffusion are coupled. The mean square displacement (MSD) of sliding shows a metastable plateau in a certain time range before it approaches the final equilibrium value. The longest sliding relaxation time scales as $\alpha^{-1/2}$, where $\alpha$ is the dimensionless moment of inertia of the rod. The rotational relaxation time obtained from the orientational correlation function is longer than that of a rod with its center fixed but faster than a rod with one end fixed. These results may be useful in understanding the dynamics of polymers connected by sliding rings.

\end{abstract}

\maketitle

\section{Introduction} 

The rotaxane\cite{doi:10.1021/ja00998a052}, which consists of a molecular chain threading through a molecular ring (e.g., crown ethers or cyclodextrins), serves as a basic unit of both biological and synthetic molecular machines\cite{https://doi.org/10.1002/anie.201703216}. The topological constraint of chains is the key feature of such mechanical unit at the nanoscale, which also introduces a unique type of bonding known as mechanical bonds in materials science\cite{doi:10.1021/acs.macromol.4c02021}. Based on this,  the slide-ring gels (SRGs)\cite{doi:10.1021/ma3021135, 2012_Ito} and the mechanically interlocked polymers (MIPs)\cite{doi:10.1021/ja0725100,doi:10.1126/science.aap7675} have been synthesized as a class of novel materials in recent decades.

The concept of topological constraints was introduced over fifty years ago alongside the idea of entanglements in long-chain polymers. The Doi-Edwards theory established a cornerstone for understanding polymer dynamics through the ``reptation'' of a chain within a confining tube \cite{1986_Doi}. The validity of this model has been shown by extensive computer simulations by Kremer and Grest\cite{10.1063/1.458541,PhysRevLett.61.566}. This tube, formed by surrounding chains, provides the physical picture of topological constraints in a highly entangled state. With the synthesis of rotaxanes and polyrotaxanes\cite{PolyrotaxaneGel}, localized entanglement was introduced and real molecular rings can slide along polymer backbones. Yasuda et al.\ observed the sliding event at the structural-unit scale by combining the quasi‑elastic neutron scattering (QENS) experiments with the full-atomistic molecular dynamics (MD) simulations\cite{doi:10.1021/jacs.9b03792}. The MD simulations have been widely employed to investigate the sliding dynamics \cite{doi:10.1021/acs.macromol.5c02674}. However, theoretical models remain limited due to the complexity of multi-chain interaction via sliding rings.

For the equilibrium statistics of sliding-chain systems, de Gennes incorporated entropic contributions into the free energy using Gaussian chain statistics\cite{DEGENNES1999231}. Subsequently, Baulin et al.\ performed a detailed analysis of the Green’s function for a Gaussian chain threading through fixed rings, providing insight into the structure of slidable grafted polymer layers\cite{doi:10.1021/ma047786w}. When multiple mobile rings are threaded onto a Gaussian chain and confined to one side of a control ring, Pinson et al.\ demonstrated that stretching the polymer produces a yield force arising from translational entropy\cite{doi:10.1021/ma4000094}. Mao et al.\ later confirmed this yielding behavior through MD simulations and refined Pinson’s free-energy model by including the excluded-volume effects of the sliding rings\cite{doi:10.1021/acs.macromol.3c02606}.

For the dynamics of sliding-chain systems, Vernerey and Lamont adapted Pinson’s free-energy model to formulate an equation of motion for single-chain dynamics. They further constructed a constitutive relation using transient network theory\cite{VERNEREY2021104212}, which successfully captured stress relaxation in SRGs. However, their model did not account for the diffusive behavior of the sliding chain. Xiong and Yu analyzed the sliding dynamics of a side chain along a fixed rod using a bead‑spring approach, observing a slow mode in segmental diffusion attributed to sliding motion\cite{2023_Xiong}. Nevertheless, the coupling between side‑chain sliding and rod rotation remains unaddressed.

Understanding the sliding dynamics of polymers is essential for designing and applying rotaxane‑based materials (e.g., molecular machines\cite{https://doi.org/10.1002/anie.201703216}, SRGs\cite{doi:10.1021/ma3021135, 2012_Ito} and MIPs\cite{doi:10.1021/ja0725100,doi:10.1126/science.aap7675}). It also provides a theoretical foundation for linking microscopic dynamics to macroscopic mechanical behavior. However, handling the localized topological constraints between molecular chains remains a challenge. In particular, the dynamical coupling between sliding and rotational diffusion in rotaxane molecules is still underdeveloped and constitutes the main focus of this work. To simplify the modeling, we study a model of the sliding dynamics in a basic rotaxane unit. The polymer backbone is modeled as a rigid rod rather than a flexible chain to limit the number of degrees of freedom. Solving the Smoluchowski equation, we show that sliding and rotation mutually influence each other: when the rod slides through the ring, an energy barrier arising from rotation-sliding coupling must be overcome. Meanwhile, when the rod rotates, sliding partially releases the constraint imposed by the fixed ring, and allows faster relaxation.

The paper is organized as follows. Section \ref{Sec_EquilibriumDistribution} derives the equilibrium distribution of rod configurations from the Hamiltonian using statistical mechanics. Section \ref{Sec_DynamicEquation} employs the Onsager variational principle, incorporating the potential energy and energy dissipation, to obtain the Smoluchowski equation. Section \ref{Sec_TimeEvolution} solves the time evolution of the distribution from the Smoluchowski equation via eigenfunction expansion. Section \ref{Sec_MSD} calculates the mean square displacement of sliding to characterize its diffusive behavior. Finally, Section \ref{Sec_RotationalRelaxation} formulates the rotational relaxation to elucidate the effect of sliding on rotation.

\section{Formulation}

\begin{figure}[htpb]
\centering
\includegraphics[width=0.46\textwidth]{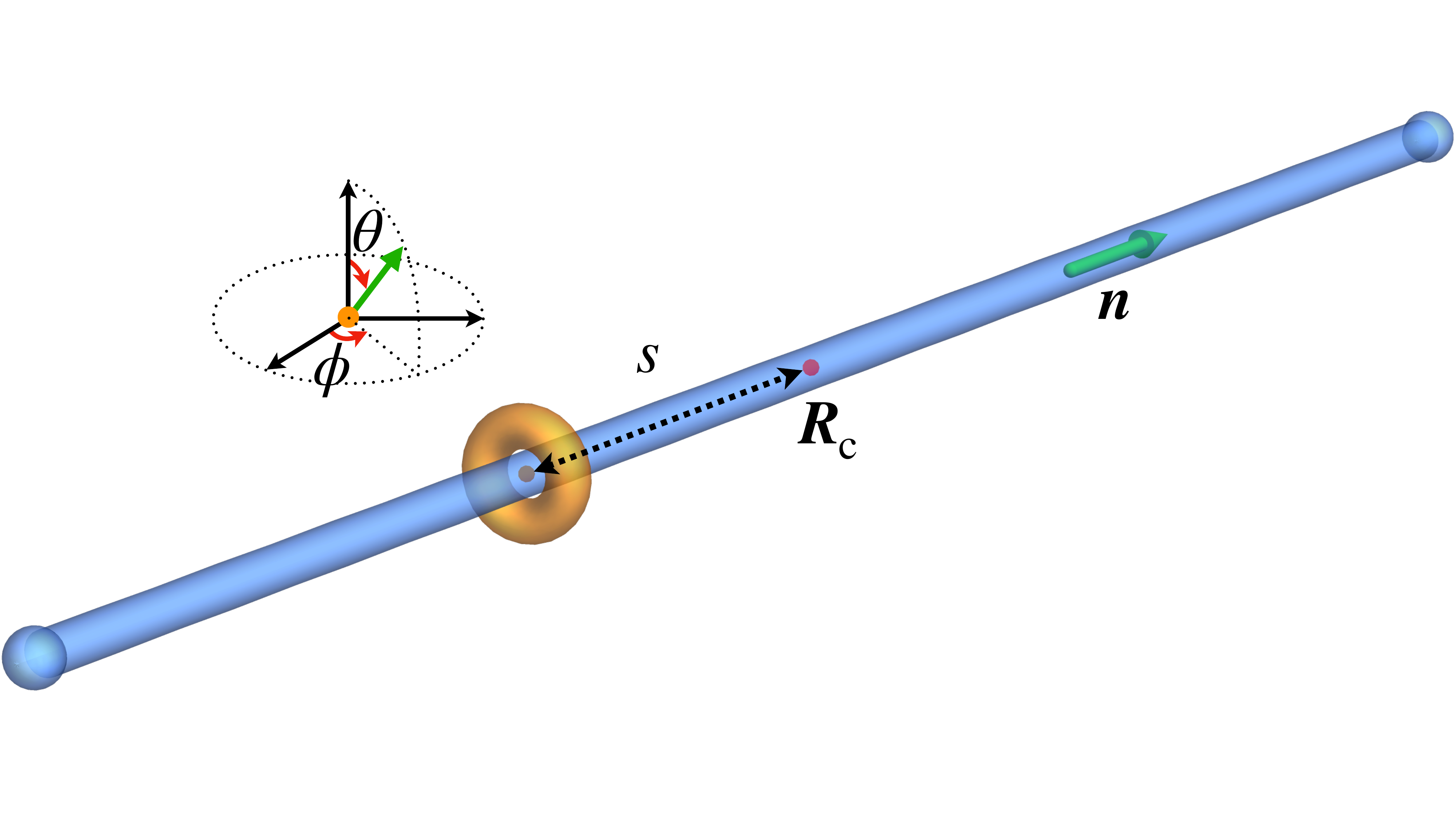}
\caption{\label{fig1}
Illustration of a sliding rod of length $L$, where $\bm{R}_\mathrm{c}$ denotes the rod-center (also the center-of-mass of the rod). The position $\bm{R}_\mathrm{c}$ is represented as $\bm{R}_\mathrm{c}=s \bm{n}$, where $s$ is the distance (which can be positive or negative) between the rod-center and the ring, and $\bm{n}=(\sin\theta\cos\phi,\sin\theta\sin\phi,\cos\theta)$ is the unit vector along the axis.}
\end{figure}

We consider a rigid rod of length $L$ (modeled as a thin, long cylinder) threading through a massless and frictionless ring whose center is fixed at the origin although it can freely rotate, as shown in Fig.~\ref{fig1}. To emphasize the role of topological constraints and to avoid dealing with the coupling effects of both sliding-translation and sliding-rotation simultaneously, the ring's position is constrained to simplify the problem. We assume that the mass distribution is not uniform and can vary along the rod axis, but the distribution is an even function with respect to the rod-center. Thus, the rod-center is also the position of rod's center-of-mass. To prevent the rod from sliding out of the ring, two stoppers are placed at both ends. The rod undergoes Brownian motion in a quiescent Newtonian fluid where sliding and rotational motions are coupled.

The configuration of the rod is described by three variables $s,\theta,\phi$ (see Fig.~\ref{fig1}): one degree of freedom for the sliding displacement $s$ of the rod-center relative to the origin, and two degrees of freedom for the unit vector along its axis $\bm{n}=(\sin\theta\cos\phi,\sin\theta\sin\phi,\cos\theta)$. Using these variables, the rod-center is given by
\begin{align}
	\bm{R}_\mathrm{c}=s\bm{n}. \label{constraint}
\end{align}
The effect of the ring is modeled as a geometric constraint, and there is no interaction potential between the rod and the ring. We also ignore the contact friction and the hydrodynamic interactions between the rod and the ring.

\subsection{Equilibrium Distribution of Rod Configurations} \label{Sec_EquilibriumDistribution}

Since there is no external potential, the Lagrangian of the system consists solely of the kinetic energy of the rod
\begin{align}
	K &= \frac{1}{2}M\dot{\bm{R}}_\mathrm{c}^2+\frac{1}{2}I\dot{\bm{n}}^2 \notag\\
	&= \frac{1}{2}\left(I+Ms^2\right)\left(\dot{\theta}^2+\dot{\phi}^2\sin^2\theta\right)+\frac{1}{2}M\dot{s}^2, \label{Ktotal}
\end{align}
where the dot is a time derivative, such as $\dot{\bm{n}}=\partial \bm{n}/\partial t$, $M$ is the total mass of the rod, and $I$ is its moment of inertia about the rod-center. If the moment of inertia of the rotating ring cannot be neglected, its contribution should also be incorporated into $I$. By using the constraint in Eq.~\eqref{constraint}, the total kinetic energy can be expressed in terms of $s$, $\theta$, and $\phi$, as shown in the second equality of Eq.~\eqref{Ktotal}. The Hamiltonian of the system in terms of the generalized momenta $p_{s}=\partial K/\partial\dot{s}$, $p_{\theta}=\partial K/\partial\dot{\theta}$, and $p_{\phi}=\partial K/\partial\dot{\phi}$ from Eq.~\eqref{Ktotal} is obtained as
\begin{equation}
	\mathcal{H} = \frac{p_{s}^2}{2M}+\frac{p_{\theta}^2}{2\left(I+Ms^2\right)}+\frac{p_{\phi}^2}{2\left(I+Ms^2\right)\sin^2\theta}. \label{Hamiltonian}
\end{equation} 

The equilibrium distribution of the rod in the phase space is proportional to $e^{-\mathcal{H}/k_\mathrm{B}T}$, where $k_\mathrm{B}$ is the Boltzmann constant and $T$ is the temperature. The equilibrium distribution of the rod configurations is obtained by integrating the Boltzmann factor over the momentum degrees of freedom\cite{doi:10.1073/pnas.71.8.3050}, i.e.,
\begin{align}
	\psi_\mathrm{eq}(s,\theta,\phi) &= z^{-1}\int dp_{s}dp_{\theta}dp_{\phi}\,e^{-\mathcal{H}(p_s,p_{\theta},p_{\phi},s,\theta,\phi)/k_\mathrm{B}T},
\end{align}
where $z = \int dsd\theta d\phi \int dp_{s}dp_{\theta}dp_{\phi}\,e^{-\mathcal{H}/k_\mathrm{B}T}$ is the partition function. The integration ranges are $p_{s},p_{\theta},p_{\phi}\in(-\infty,\infty)$, $s\in[-L/2,L/2]$, $\theta\in[0,\pi]$, and $\phi\in[0,2\pi]$. 

By using the Hamiltonian in Eq.~\eqref{Hamiltonian}, the equilibrium distribution of the sliding rod is given by
\begin{align}
	\psi_\mathrm{eq}(s,\theta,\phi) = \frac{3\left( \alpha L^2+s^2\right)\sin \theta}{(12\alpha+1)\pi L^3},
\end{align}
where $I=\alpha ML^2$ has been used, and $\alpha$ is given by
\begin{align}
	\alpha = \left(\int_{-1/2}^{1/2}d\tilde{s}\,\rho(\tilde{s})\right)^{-1}\int_{-1/2}^{1/2}d\tilde{s}\,\rho(\tilde{s})\tilde{s}^2,
\end{align}
with $\tilde{s}=s/L$. The function $\rho(\tilde{s})$ is the mass distribution, which is an even function. The parameter $\alpha$ characterizes different distributions and changes from $0$ to $1/4$. For instance, $\alpha\to 0$ indicates that the mass is concentrated at the rod-center, $\alpha=1/12$ corresponds to a uniform mass distribution along the rod, and $\alpha=1/4$ represents mass concentrated at both ends. Since the probability $\psi_\mathrm{eq}(s,\theta,\phi) ds d\theta d\phi = \psi_\mathrm{eq}(s,\bm{n}) d\Omega$, where $d\Omega= d\bm{n}ds=\sin\theta d\theta d\phi ds$ denotes the volume element in the configuration space, the equilibrium distribution in terms of $s$ and $\bm{n}$ becomes
\begin{equation}
	\psi_\mathrm{eq}(s,\bm{n}) = \frac{3\left( \alpha L^2+s^2\right)}{(12\alpha+1)\pi L^3}. \label{equilibrium}
\end{equation}

Equation~\eqref{equilibrium} shows that the equilibrium distribution is uniform in orientation but quadratic in the sliding distance $s$. A similar expression was obtained in a study of a rod escaping through a hole \cite{doi:10.1021/acs.jpcb.2c03976}, reflecting the coupling between rotational and sliding motions. The nonuniform $\psi_\mathrm{eq}$ is a purely entropic effect induced by the constraint Eq.~\eqref{constraint} and from the coordinate-dependent moment of inertia (generally metric determinant\cite{doi:10.1073/pnas.71.8.3050}), not by an external potential. If the orientation $\bm{n}$ is fixed, the distribution of rod configurations becomes uniform. In the presence of orientation, the distribution function has a minimum at $s=0$. This minimum approaches zero when mass is concentrated near the rod-center, indicating that the rod tends to slide away from the central position.

\subsection{Dynamic Equation of a Sliding Rod} \label{Sec_DynamicEquation}

We now derive the time evolution equation for the distribution function $\psi(s,\bm{n},t)$ using the Onsager variational principle described in Ref \cite{doi2013soft}. Let $\psi(s,\bm{n},t)$ be the distribution function for the rod in the configuration $(s, \bm{n})$ at time $t$. This distribution function satisfies the continuity equation
\begin{align}
	\dot{\psi} = -\frac{\partial }{\partial s}\left(\dot{s}\psi\right) -\frac{\partial}{\partial \bm{n}}\cdot\left(\dot{\bm{n}}\psi\right).\label{conservation}
\end{align}

According to the Onsager variational principle, the velocities $\dot{s}$ and $\dot{\bm{n}}$ in Eq.~\eqref{conservation} are determined by minimizing the Rayleighian $\mathcal{R}$. Additionally, the unit vector constraint $\bm{n}\cdot\dot{\bm{n}}=0$ is incorporated into the Rayleighian using a Lagrange multiplier $\mu$ (generally depends on $s$ and $\bm{n}$). Therefore, the Rayleighian of the system is
\begin{align}
	\mathcal{R}[\dot{s},\dot{\bm{n}}] &= \dot{A}[\dot{s},\dot{\bm{n}}]+\Phi[\dot{s},\dot{\bm{n}}]+\int d\Omega\,\mu\bm{n}\cdot\dot{\bm{n}}, \label{Rayleighian}
\end{align} 
where $\dot{A}$ is the change rate of free energy and $\Phi$ is the dissipation function which we shall describe below.

\paragraph{Free Energy.} Because the distribution of the sliding distance $s$ is non-uniform, the rod experiences an effective potential energy $U(s)$ when sliding through the ring. The equilibrium distribution in Eq.~\eqref{equilibrium} can be recast into $\psi_\mathrm{eq}(s,\bm{n})=z_0^{-1}e^{-U(s)/k_\mathrm{B}T}$, where $z_0$ is the normalization constant. Thus, the effective potential energy is given by
\begin{align}
	U(s) = -k_\mathrm{B}T\ln\left(1+\frac{s^2}{\alpha L^2}\right),
\end{align}
if one chooses a reference level as $U(0)=0$.

Therefore, the free energy change rate of the sliding rod is
\begin{align}
	\dot{A} &= \frac{\partial}{\partial t}\int d\Omega\,\psi\left(U+k_\mathrm{B} T\ln\psi\right) \notag\\
	&= k_\mathrm{B} T\int d\Omega\,\psi\left(\dot{\bm{n}}\cdot\frac{\partial \ln\psi}{\partial \bm{n}} + \dot{s}\frac{\partial \ln\psi}{\partial s} + \frac{\dot{s}}{k_\mathrm{B}T}\frac{\partial U}{\partial s}\right), \label{free_energy_change_rate}
\end{align}
where the vanishing flux condition $j=\dot{s}\psi=0$ at the boundaries $s=\pm L/2$ has been used, ensuring that the rod cannot slide out of the ring.

\paragraph{Energy Dissipation.} We ignore the friction between the ring and rod, and assume that the energy dissipation is due to the motion of the rod in a viscous fluid. Since the rod-center is moving at the rate $\dot{\bm{R}}_\mathrm{c}$ and the rod is rotating with the angular velocity $\dot{\bm{n}}$, the energy dissipation function is written as\cite{1986_Doi,Kim_1991}
\begin{align}
	\Phi &= \int d\Omega\,\psi\left(\frac{1}{2}\dot{\bm{R}}_\mathrm{c}\cdot\left[\zeta_\parallel\bm{n}\bm{n}+\zeta_\perp(\bm{\delta}-\bm{n}\bm{n})\right]\cdot\dot{\bm{R}}_\mathrm{c}+\frac{1}{2}\zeta_0\dot{\bm{n}}^2\right) \notag\\
	&= \int d\Omega\,\psi\left(\frac{1}{2}\left(\zeta_0+\zeta_\perp s^2\right)\dot{\bm{n}}^2+\frac{1}{2}\zeta_\parallel\dot{s}^2\right), \label{dissipation_function}
\end{align}
where $\zeta_\parallel$ and $\zeta_\perp$ are the translational friction coefficients parallel and perpendicular to the axis, respectively, and $\zeta_0$ is the rotational friction coefficient about the rod-center (i.e., about $s=0$). Using Eq.~\eqref{constraint}, we obtain the second equality, in which the last term represents the dissipation due to sliding when the ring is located at $s$, whereas the first term corresponds to the dissipation associated with the rotation of the rod around $s$, where the rotational friction coefficient satisfies an analogy of the parallel axis theorem.

\paragraph{Smoluchowski Equation.} Substituting $\dot{A}$ and $\Phi$ in Eqs.~\eqref{free_energy_change_rate} and \eqref{dissipation_function} into Eq.~\eqref{Rayleighian}, we obtain the explicit expression for the Rayleighian. The minimization conditions $\delta\mathcal{R}/\delta\dot{s}=0$ and $\delta\mathcal{R}/\delta\dot{\bm{n}}=0$ yield
\begin{subequations}\label{solutions}
\begin{align} 
    \dot{s} &= -\frac{k_\mathrm{B} T}{\zeta_\parallel}\left(\frac{\partial \ln\psi}{\partial s}+\frac{1}{k_\mathrm{B}T}\frac{\partial U}{\partial s}\right), \\
    \dot{\bm{n}} &= -\frac{k_\mathrm{B} T}{\zeta_0+\zeta_\perp s^2} 
    \left(\bm{\delta}-\bm{n}\bm{n}\right)\cdot\frac{\partial \ln\psi}{\partial\bm{n}},
\end{align}
\end{subequations}
where the Lagrange multiplier has been determined from the condition $\bm{n}\cdot\dot{\bm{n}}=0$, and is given by $\mu=-k_\mathrm{B} T\bm{n}\cdot\partial \psi/\partial \bm{n}$.

Combining Eqs.~\eqref{conservation} and \eqref{solutions} gives the Smoluchowski equation
\begin{align}
	\frac{\partial \psi}{\partial t} &= D_\parallel\frac{\partial }{\partial s}\left(\frac{\partial \psi}{\partial s}+\frac{\psi}{k_\mathrm{B}T}\frac{\partial U}{\partial s}\right) \notag\\
	&\quad+ D_0\frac{1}{1+\beta_\perp s^2/L^2}\frac{\partial }{\partial \bm{n}}\cdot\left(\bm{\delta}-\bm{n}\bm{n}\right)
    \cdot\frac{\partial}{\partial \bm{n}}\psi, 
    \label{Smoluchowski}
\end{align}
where $D_\parallel = k_\mathrm{B} T/\zeta_\parallel$ and $D_0 = k_\mathrm{B} T/\zeta_0$ are the (parallel) translational and rotational diffusion coefficient, respectively. According to the slender-body theory\cite{1986_Doi,Kim_1991}, $\beta_\perp = \zeta_\perp L^2/\zeta_0$ and $\beta_\parallel = \zeta_\parallel L^2/\zeta_0$ are generally dependent on the particle shape or the aspect ratio of the rod. For a thin and long rod, we approximately have $\beta_\perp \approx 12$ and $\beta_\parallel \approx 6$.

\section{Sliding diffusion and rotational relaxation of rod}

We first solve the Smoluchowski equation via eigenfunction expansion. By introducing the dimensionless variables $\tilde{s}=s/L$, $\tilde{t}=D_0t$, and $\tilde{\psi}=\psi L$, Eq.~\eqref{Smoluchowski} can be written in the dimensionless form as
\begin{align}
	\frac{\partial \tilde{\psi}}{\partial \tilde{t}} &= \frac{1}{\beta_\parallel}\frac{\partial }{\partial \tilde{s}}\left(\frac{\partial \tilde{\psi}}{\partial \tilde{s}}-\frac{2\tilde{s}}{\alpha+\tilde{s}^2}\tilde{\psi}\right) \notag\\
	&\quad+ \frac{1}{1+\beta_\perp \tilde{s}^2} \left(\frac{1}{\sin\theta}\frac{\partial}{\partial \theta}\sin\theta\frac{\partial }{\partial \theta}+\frac{1}{\sin^2\theta}\frac{\partial^2 }{\partial \phi^2}\right)\tilde{\psi}, \label{smleq}
\end{align}
with the normalization condition
\begin{align}
	\int_{-1/2}^{1/2}d\tilde{s} \int_0^{2\pi}d\phi \int_0^\pi d\theta\,\sin\theta \tilde{\psi} (\tilde{s},\theta,\phi,\tilde{t}) = 1.
\end{align}
The boundary conditions are $[\partial \tilde{\psi}/\partial \tilde{s}-2\tilde{s}\tilde{\psi}/(\alpha+\tilde{s}^2)]_{\tilde{s}=\pm 1/2}=0$ (i.e., the flux vanishes at the boundaries $j(\tilde{s})|_{\tilde{s}=\pm 1/2}=0$ ensuring that the rod cannot slide out of the ring), the periodic boundary condition $\tilde{\psi}|_{\phi=0}=\tilde{\psi}|_{\phi=2\pi}$, and the boundedness conditions of $\tilde{\psi}$ at $\theta=0,\pi$. Then, the equilibrium distribution can be derived as 
\begin{align}
	\tilde{\psi}_{\mathrm{eq}}(\tilde{s},\theta,\phi) = \frac{3}{(12\alpha+1)\pi}\left( \alpha+\tilde{s}^2\right), \label{equilibrium_dimensionless}
\end{align}
which coincides with Eq.~\eqref{equilibrium}.

Assuming that the initial condition is given by
\begin{align}
	\left.\tilde{\psi} (\tilde{s},\theta,\phi,\tilde{t})\right|_{\tilde{t}=0}=\frac{1}{\sin\theta}\delta(\tilde{s}-\tilde{s}')\delta(\theta-\theta')\delta(\phi-\phi'), \label{initial}
\end{align}
where $\delta(x)$ is the Dirac delta function, and $\tilde{s}',\theta',\phi'$ are values at time $\tilde{t}=0$, we obtain the solution of Eq.~\eqref{smleq} for the Green's function (denoted by $\mathcal{G}$). Performing the eigenfunction expansion (details are shown in Appendix~\ref{app:A}) yields
\begin{widetext}
\begin{equation} \label{Green'sFunction}
	\mathcal{G}(\tilde{s},\theta,\phi,\tilde{t};\tilde{s}',\theta',\phi') = \tilde{\psi}_{\mathrm{eq}}^{1/2}(\tilde{s})\tilde{\psi}_{\mathrm{eq}}^{-1/2}(\tilde{s}')\sum_{p=0}^{g}\sum_{n=0}^{g}\sum_{q=0}^{g}\sum_{l=0}^\infty\sum_{m=-l}^l a_n^{lp}a_q^{lp}\varphi_n(\tilde{s})\varphi_{q}(\tilde{s}')Y_l^{m} (\theta,\phi)Y_{l}^{m} (\theta',\phi') e^{-\lambda_{lp}\tilde{t}},
\end{equation}
\end{widetext}
where $g$ is a truncation number, set to be here $g=20$ (which is sufficient to obtain numerically converged solutions compared with those for $g=30$). In the above, $\lambda_{lp}$ and $a_n^{lp}$ are the eigenvalues and eigenvectors of the matrix~\eqref{MM}, respectively, $\varphi_n(\tilde{s})$ are the eigenfunctions of Eq.~\eqref{eigenvalue_problem}, and $Y_l^{m} (\theta,\phi)$ are the spherical harmonics. When $l=0$ and $p=0$ we have $\lambda_{00} = 0$ and the corresponding eigenfunction $\varphi_0$ (see Eq.~\eqref{eigenfunction_0}) and eigenvector $a_n^{0p}$ (see Eq.~\eqref{eigenvector_0}).

The autocorrelation function of a quantity $\mathcal{F}(\tilde{s},\theta,\phi)$ is generally obtained from the Green's function and the initial equilibrium distribution $\tilde{\psi}_\mathrm{eq}$ at $\tilde{t}=0$ by
\begin{align}
 	\langle \mathcal{F}(\tilde{t})\mathcal{F}(0)\rangle &= \int d\tilde{\Omega} \int d\tilde{\Omega}'\,\mathcal{G}(\tilde{\Omega},\tilde{t};\tilde{\Omega}')\tilde{\psi}_\mathrm{eq}(\tilde{\Omega}')\mathcal{F}(\tilde{\Omega})\mathcal{F}(\tilde{\Omega}') \notag\\
 	&= \sum_{p=0}^{g}\sum_{l=0}^\infty\sum_{m=-l}^l  G_{lpm}^2e^{-\lambda_{lp}\tilde{t}},  \label{EnsembAverage}
\end{align}	
where $\tilde{\Omega}=(\tilde{s},\theta,\phi)$ and $d\tilde{\Omega}=\sin\theta d\theta d\phi d\tilde{s}$ for shorthand, and
\begin{align}
	G_{lpm} &= \sum_{n=0}^{g}a_n^{lp}\int_{-1/2}^{1/2}d\tilde{s}\,\tilde{\psi}_{\mathrm{eq}}^{1/2}(\tilde{s})\varphi_n(\tilde{s}) \notag\\
	&\quad\times\int_0^{2\pi}d\phi\int_0^\pi d\theta\, \sin\theta Y_l^{m} (\theta,\phi)\mathcal{F}(\tilde{s},\theta,\phi).
\end{align}

\subsection{Time Evolution of Distribution toward Equilibrium} \label{Sec_TimeEvolution}

The time evolution of the distribution $\tilde{\psi} (\tilde{s},\tilde{t})$ is defined as
\begin{align}
	\tilde{\psi} (\tilde{s},\tilde{t}) &= \int_0^{2\pi}d\phi\int_0^\pi d\theta \,\sin\theta\tilde{\psi} (\tilde{s},\theta,\phi,\tilde{t}) \notag\\
	&= \tilde{\psi}_{\mathrm{eq}}^{1/2}(\tilde{s})\sum_{p=0}^{g}c_p \varphi_p(\tilde{s})e^{-\lambda_{0p}\tilde{t}}, \label{psi_s}
\end{align}
with
\begin{align}
	c_p  = \int_{-1/2}^{1/2}d\tilde{s}\,\tilde{\psi}_{\mathrm{eq}}^{-1/2}(\tilde{s})\varphi_p(\tilde{s})\tilde{\psi}_\mathrm{in}(\tilde{s}),
\end{align}
where Eq.~\eqref{solution_expansion} and \eqref{coefficients_A} have been used to obtain the second equality in Eq.~\eqref{psi_s}. The initial distribution is given by $\tilde{\psi}_\mathrm{in}(\tilde{s})$ at time $\tilde{t}=0$ to determine the coefficients $c_p$.

\begin{figure*}[htpb]
\centering
\includegraphics[width=0.32\textwidth]{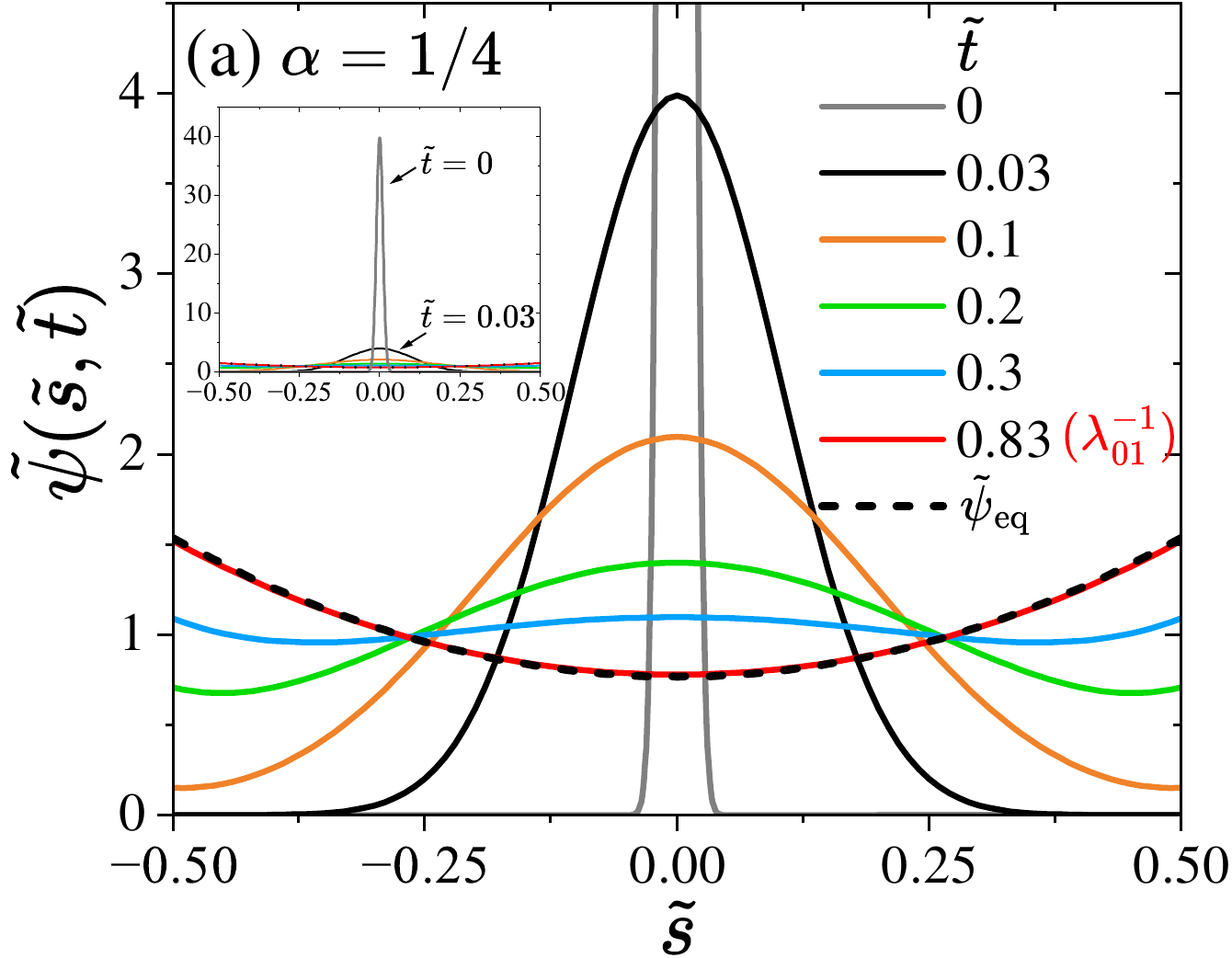}
\includegraphics[width=0.32\textwidth]{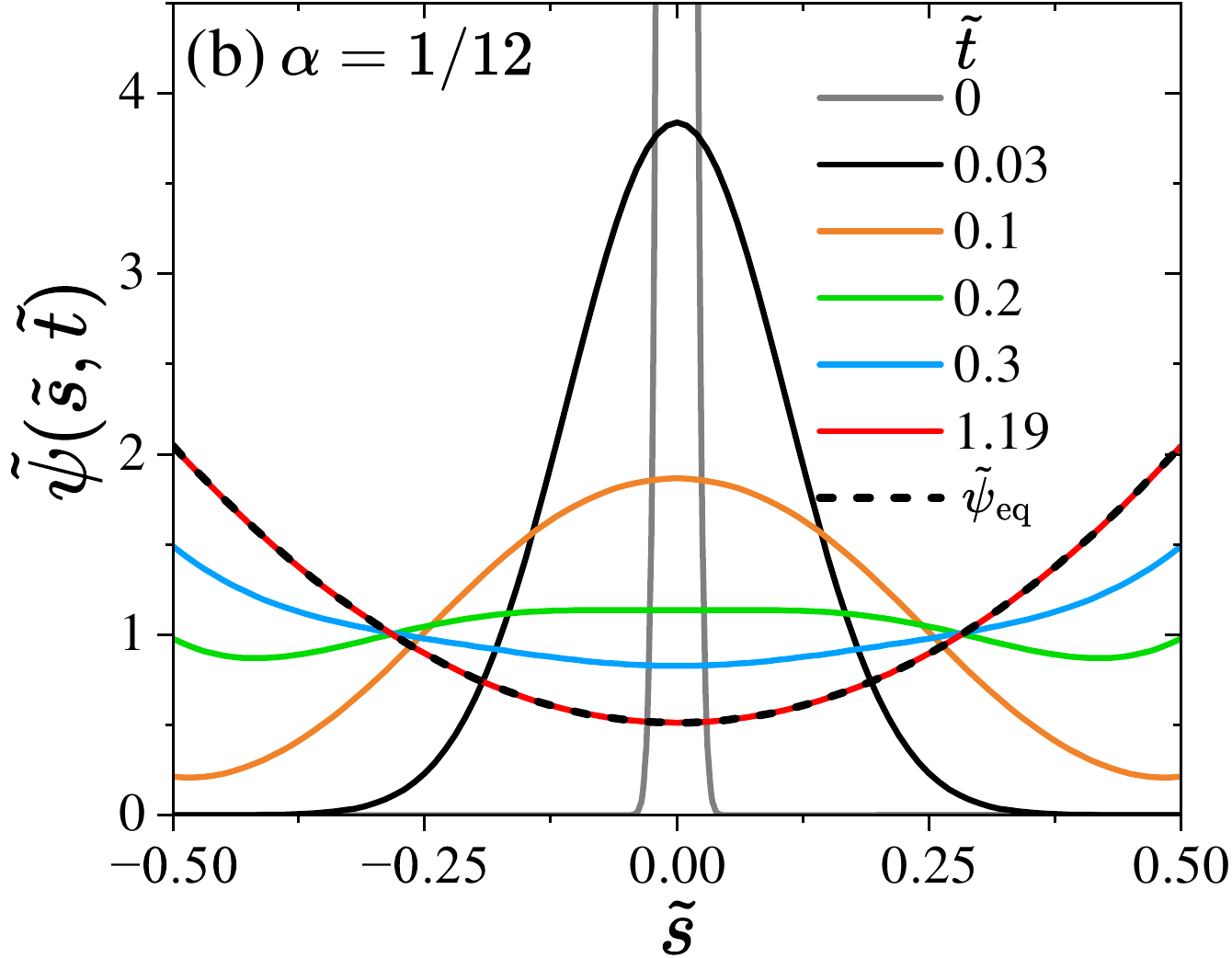}
\includegraphics[width=0.32\textwidth]{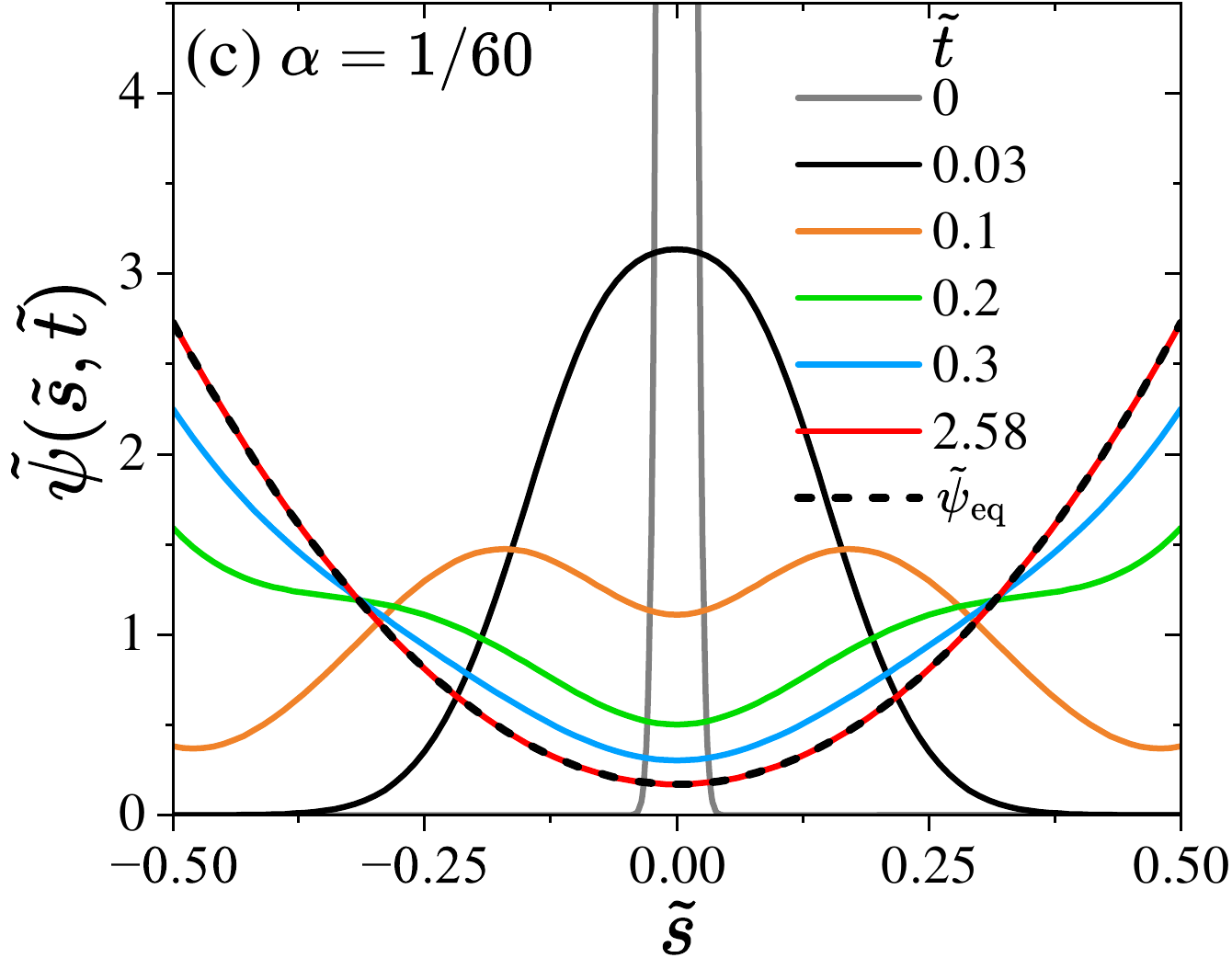}
\caption{\label{fig2}
Time evolution of the probability distribution of sliding distance $\tilde{\psi} (\tilde{s},\tilde{t})$ under different mass distributions ($\beta_\perp = 12$ and $\beta_\parallel = 6$ have been used): (a) $\alpha=1/4$ (mass concentrated at both ends), (b) $\alpha=1/12$ (uniform distribution along the rod), and (c) $\alpha=1/60$ (mass concentrated at the rod-center). The initial condition $\tilde{\psi}_\mathrm{in}(\tilde{s})$ at $\tilde{t}=0$ is a Gaussian distribution with $\mu=0$ and $\sigma^2=0.0001$ (see inset in panel (a)). Red solid lines show the distributions at the longest relaxation time $\lambda_{01}^{-1}$, and black dashed lines represent the equilibrium distributions $\tilde{\psi}_{\mathrm{eq}}(\tilde{s})$.}
\end{figure*}

Here, the initial distribution $\tilde{\psi}_\mathrm{in}(\tilde{s})$ is set to be a Gaussian distribution $\mathcal{N}(\mu,\sigma^2)$ with the mean value $\mu=0$ and the variance $\sigma^2=0.0001$ (see the insert in Fig.~\ref{fig2}(a) at $\tilde{t}=0$). Figure~\ref{fig2} shows the time evolution of the probability distribution toward equilibrium for different mass distributions $\alpha$. When the mass is concentrated at the two ends of the rod ($\alpha=1/4$, Fig.~\ref{fig2}(a)), the rod slides away from the center rapidly and reaches the equilibrium within a timescale $\lambda_{01}^{-1}$. A qualitatively similar relaxation occurs for a uniform mass distribution ($\alpha=1/12$, Fig.~\ref{fig2}(b)) and for mass concentrated at the rod-center ($\alpha=1/60$, Fig.~\ref{fig2}(c)). However, the transient behavior differs among these cases. Notably, a double-peaked distribution appears when the mass is concentrated at the center as in Fig.~\ref{fig2}(c). This is because the energy barrier at $s=0$ is higher than the other cases, and the driving force to slide away from the center is larger.

\subsection{Mean Square Displacement of Sliding} \label{Sec_MSD}

We now examine the Brownian sliding motion. If the initial condition is set to the equilibrium distribution in Eq.~\eqref{equilibrium_dimensionless}, it follows that $\langle \tilde{s}^2(\tilde{t}) \rangle=\langle \tilde{s}^2(0) \rangle=\langle \tilde{s}^2\rangle_\mathrm{eq}$ with
\begin{align}
	\langle \tilde{s}^2\rangle_\mathrm{eq} = \frac{20 \alpha +3}{20(12 \alpha +1)}. \label{equilibrium_s2}
\end{align}
Therefore, the mean square displacement (MSD) of sliding is
\begin{align}
	&\langle [\tilde{s}(\tilde{t})-\tilde{s}(0)]^2 \rangle = 2\langle \tilde{s}^2\rangle_\mathrm{eq}-2\langle\tilde{s}(\tilde{t})\tilde{s}(0) \rangle \notag\\
	&= \frac{20 \alpha +3}{10(12 \alpha +1)}-2\sum_{p=1}^{g}\left(\int_{-1/2}^{1/2}d\tilde{s}\,\varphi_0(\tilde{s})\varphi_p(\tilde{s})\tilde{s}\right)^2e^{-\lambda_{0p}\tilde{t}},
\end{align}
where $\langle \tilde{s}(\tilde{t})\tilde{s}(0)\rangle$ is obtained from Eq.~\eqref{EnsembAverage} by setting $\mathcal{F}=\tilde{s}$. 

\begin{figure*}[htpb]
\centering
\includegraphics[width=1\textwidth]{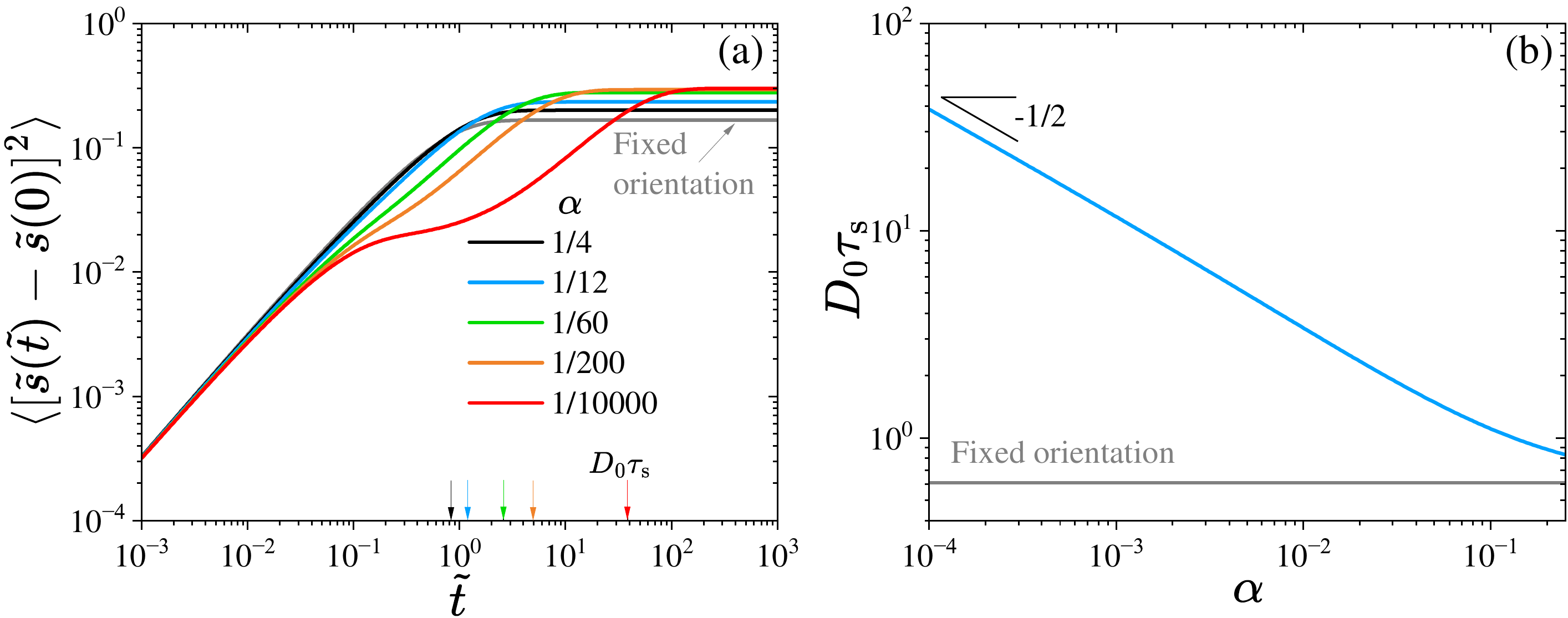}
\caption{\label{fig3}
(a) Mean square displacement (MSD) of the sliding distance for different mass distribution parameters $\alpha$. For comparison, the MSD for a fixed rod orientation is also shown, denoted by ``Fixed orientation''. (b) Longest sliding relaxation time $\tau_\mathrm{s}$ (scaled by $D_0$) is plotted against $\alpha$. $\tau_\mathrm{s}$ decreases with the increase of $\alpha$. For a rod of fixed orientation, $\tau_\mathrm{s}$ is independent of $\alpha$ and is shown by the black line. $\beta_\perp = 12$ and $\beta_\parallel = 6$ have been used.}
\end{figure*}

The transient behavior of MSD is shown in Fig.~\ref{fig3}(a). The MSD eventually reaches the equilibrium value since the rod cannot slide out of the ring. For short time $D_0t\ll1$, it can be shown that the short-time sliding diffusion coefficient is
\begin{align}
	D_\mathrm{s} = \frac{1}{2t}\langle [s(t)-s(0)]^2 \rangle \approx D_\parallel. \label{sliding_diffusion}
\end{align}
The last expression is obtained by performing Taylor expansion of the MSD for short time.

For small $\alpha$, the MSD shows a plateau at short time (see the red line in Fig.~\ref{fig3}(a)). It can be justified as follows: When the mass is concentrated at the rod-center ($\alpha\to 0$), the energy barrier becomes markedly higher. Consequently, within short time intervals, the probability of crossing the energy barrier is small. This leads to a metastable confinement within a half side of the rod, described by the distribution function $\tilde{\psi}_\mathrm{meta}(\tilde{s})=24\tilde{s}^2$ over the interval $\tilde{s}\in[0,1/2]$ (or $\tilde{s}\in[-1/2,0]$). Then, the MSD at the metastable state can be evaluated as
\begin{align}
	\langle [\tilde{s}-\langle\tilde{s}\rangle_\mathrm{meta}]^2 \rangle_\mathrm{meta} \approx 0.0094, \label{MSD_meta}
\end{align}
where $\langle \cdot \rangle_\mathrm{meta} = \int_0^{1/2}d\tilde{s}\,\tilde{\psi}_\mathrm{meta}(\tilde{s})(\cdot)$. According to Eq.~\eqref{sliding_diffusion}, we have $\langle [\tilde{s}(\tilde{t})-\tilde{s}(0)]^2 \rangle=2\tilde{t}/\beta_\parallel$ within short time intervals. Thus, the time scale (scaled by $D_0$) to reach the metastable state is evaluated as $\tilde{t}_\mathrm{meta} \approx 0.0282$ based on Eq.~\eqref{MSD_meta}. This value agrees with the short-time plateau observed in the MSD (the red line in Fig.~\ref{fig3}(a)). Therefore, the short-time plateau is a signature of the metastable state in which the rod is temporarily trapped before overcoming the barrier.

Figure~\ref{fig3}(b) shows the dependence of the longest sliding relaxation time $\tau_\mathrm{s}$ on $\alpha$. Here, $\tau_\mathrm{s} = (\lambda_{01} D_0)^{-1}$ represents the characteristic time for the MSD to approach its final equilibrium value. By using the eigenvalue equation \eqref{eigenvalue_problem}, the smallest nonvanishing eigenvalue $\lambda_{01}$ can be approximately evaluated as
\begin{align}
	\lambda_{01} = \frac{1}{\beta_\parallel}\int_{-1/2}^{1/2}d\tilde{s}\,\varphi_1\mathcal{L}\varphi_1 \sim \int_{-1/2}^{1/2}d\tilde{s}\,\frac{\alpha\tilde{s}^2}{(\alpha+\tilde{s}^2)^2} \sim \alpha^{1/2},
\end{align}
where $\varphi_1 \sim \tilde{s}$ has been used when $\alpha$ is small. The last expression is obtained by keeping the lowest order in the series of $\alpha$. Therefore, the scaling $\tau_\mathrm{s} \sim \alpha^{-1/2}$ holds when the mass is concentrated at the rod-center. Compared to a rod with fixed orientation, relaxation to the final equilibrium is delayed when rotation is allowed, due to the presence of an effective energy barrier.

\subsection{Rotational Relaxation} \label{Sec_RotationalRelaxation}

To characterize the rotational relaxation, we employ the orientational correlation function $\langle \bm{n}(\tilde{t})\cdot\bm{n}(0)\rangle$, which can be calculated from Eq.~\eqref{EnsembAverage} by setting $\mathcal{\bm{F}}=\bm{n}$. This yields
\begin{equation}
	\langle\bm{n}(\tilde{t})\cdot\bm{n}(0)\rangle = \sum_{p=0}^{g}(a_0^{1p})^2e^{-\lambda_{1p}\tilde{t}}.
\end{equation}
Note that $a_0^{1p}$ and $\lambda_{1p}$ are determined by the eigenvalue problem that depends on the sliding dynamics.

\begin{figure*}[htpb]
\centering
\includegraphics[width=1\textwidth]{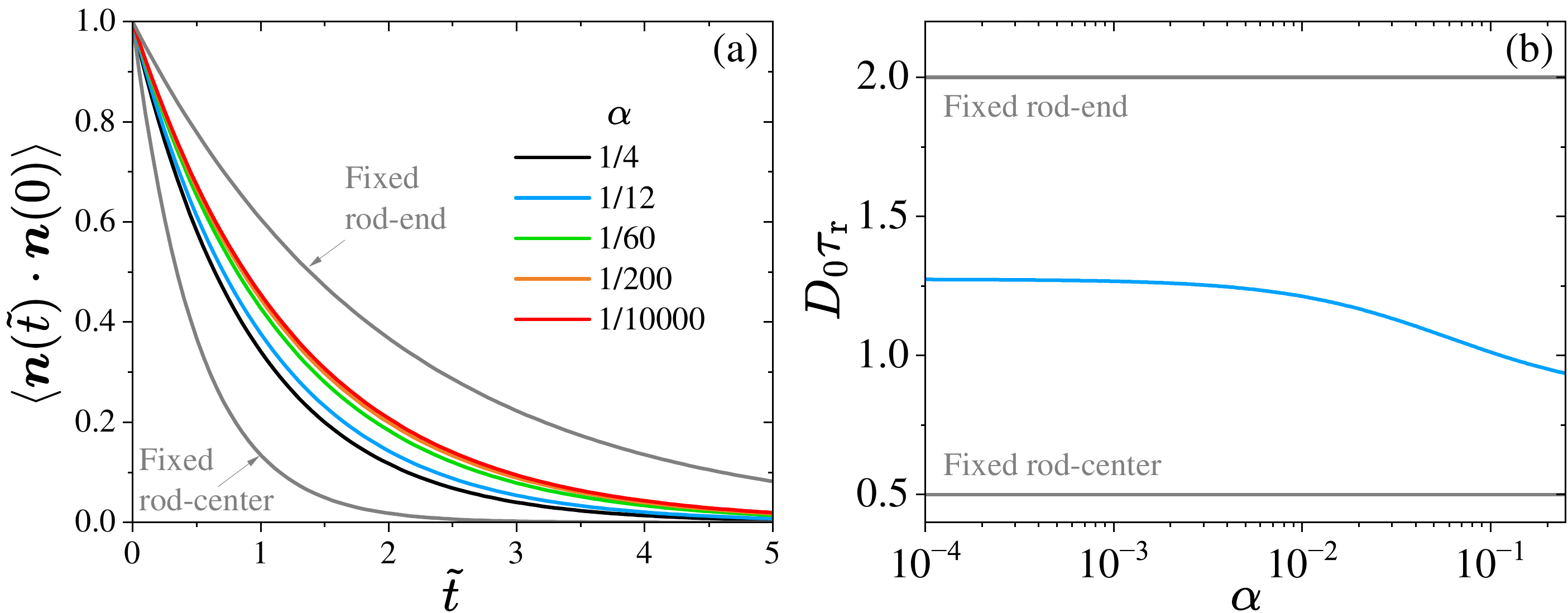}
\caption{\label{fig4}
(a) Orientational correlation function $\langle\bm{n}(\tilde{t})\cdot\bm{n}(0)\rangle$ for a sliding rod with different mass distribution parameters $\alpha$. For comparison, results for a non-sliding rod with its center fixed at the origin (``Fixed rod-center'') and with one end fixed (``Fixed rod-end'') are shown. (b) Longest rotational relaxation time $\tau_\mathrm{r}$ (scaled by $D_0$) of the sliding rod as a function of $\alpha$. The corresponding times for the fixed-center and fixed-end cases are independent of $\alpha$ and are included for comparison. $\beta_\perp = 12$ and $\beta_\parallel = 6$ have been used.}
\end{figure*}

Figure~\ref{fig4}(a) displays the transient decay of the orientational correlation of a sliding rod for various mass distributions $\alpha$. For comparison, two non-sliding cases are included; one when the rod-center is fixed at the origin, and the other when one rod-end is fixed. The rotational relaxation of the sliding rod is slower than that of the center-fixed rod, but faster than that of the end-fixed rod. Thus, although the fixed ring constrains and slows rotational relaxation, sliding partially releases this constraint and allows for faster relaxation.

From the MSD $\langle [\bm{n}(\tilde{t})-\bm{n}(0)]^2 \rangle=2[1-\langle\bm{n}(\tilde{t})\cdot\bm{n}(0)\rangle]$, the short‑time rotational diffusion constant is defined as
\begin{align}
	D_\mathrm{r} = \frac{1}{4t}\langle [\bm{n}(t)-\bm{n}(0)]^2 \rangle \approx D_0\sum_{p=0}^{g}\frac{(a_0^{1p})^2}{2}\lambda_{1p},
\end{align}
for $D_0t\ll1$. We find that $D_\mathrm{r}$ increases with increasing $\alpha$, according to the numerical calculation.

The time scale of rotational relaxation can be defined by the longest relaxation time as $\tau_\mathrm{r}=(\lambda_{10}D_0)^{-1}$, which marks the time for the rotational MSD to reach equilibrium. Figure~\ref{fig4}(b) shows the dependence of $\tau_\mathrm{r}$ on $\alpha$. The relaxation time of the sliding rod also lies between those of the two non-sliding reference cases.

\begin{figure}[htpb]
\centering
\includegraphics[width=0.5\textwidth]{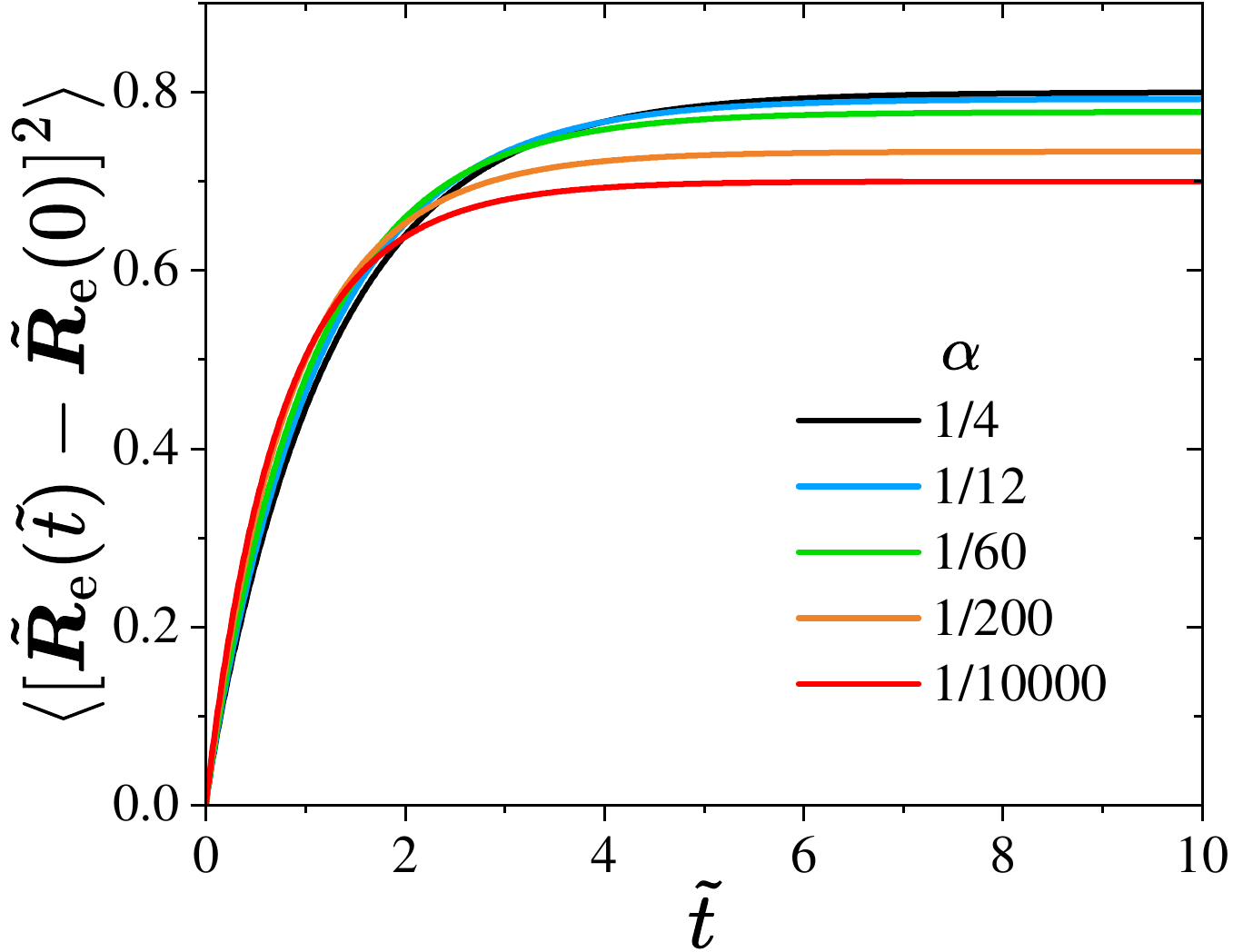}
\caption{\label{fig5}
Mean square displacement (MSD) of the rod end for different mass distributions $\alpha$, where the rod-end is defined as $\tilde{\bm{R}}_\mathrm{e}(\tilde{t}) = [\tilde{s}(\tilde{t}) + 1/2]\bm{n}(\tilde{t}) $.  $\beta_\perp = 12$ and $\beta_\parallel = 6$ have been used.}
\end{figure}

The position of a rod-end can be expressed as $\tilde{\bm{R}}_\mathrm{e}(\tilde{t}) = [\tilde{s}(\tilde{t}) +1/2]\bm{n}(\tilde{t}) $ in terms of $\tilde{s}$ and $\bm{n}$. Its MSD is therefore given by
\begin{align}
	&\langle [\tilde{\bm{R}}_\mathrm{e}(\tilde{t})-\tilde{\bm{R}}_\mathrm{e}(0)]^2 \rangle = \frac{1}{2}[1 -\langle\bm{n}(\tilde{t})\cdot\bm{n}(0)\rangle] \notag\\
	&\quad+ 2\left[\langle\tilde{s}^2\rangle_\mathrm{eq}-\langle\tilde{s}(\tilde{t})\tilde{s}(0)\bm{n}(\tilde{t})\cdot\bm{n}(0)\rangle\right] \notag\\
	&= \frac{20 \alpha +3}{10(12 \alpha +1)}+\frac{1}{2}\left(1 -\sum_{p=0}^{g}(a_0^{1p})^2e^{-\lambda_{1p}\tilde{t}}\right) \notag\\
	&\quad-2\sum_{p=0}^{g}  \left(\sum_{n=0}^{g}a_n^{1p}\int_{-1/2}^{1/2}d\tilde{s}\,\varphi_0(\tilde{s})\varphi_n(\tilde{s})\tilde{s}\right)^2e^{-\lambda_{1p}\tilde{t}},
\end{align}
where the correlation $\langle\tilde{s}(\tilde{t})\tilde{s}(0)\bm{n}(\tilde{t})\cdot\bm{n}(0)\rangle$ is calculated from Eq.~\eqref{EnsembAverage} by setting $\mathcal{F}=\tilde{s}\bm{n}$. The transient behavior is shown in Fig.~\ref{fig5}, which confirms that rotational relaxation is nearly completed within the time scale $\tau_\mathrm{r}$, reaching a final equilibrium state that depends on $\alpha$ (see Eq.~\eqref{equilibrium_s2}).

\section{Conclusion and Discussion} 

We have studied the Brownian motion of a rigid rod threading through a ring with fixed ring-center by solving the corresponding Smoluchowski equation. The equilibrium distribution of the sliding distance is non‑uniform and quadratic, arising from the sliding-rotational coupling and influenced by the rod's mass distribution through its moment of inertia. This results in an effective energy barrier for the rod to slide from one side of the rod to the other. The barrier becomes significant when mass is concentrated near the rod-center, leading to a longer relaxation time from the initial distribution to the final equilibrium distribution. Correspondingly, a metastable plateau appears in the MSD of sliding, reflecting the transient trapping before overcoming the barrier. The longest sliding relaxation time scales with the moment of inertia as $\alpha^{-1/2}$. Furthermore, the rotational relaxation of the sliding rod is slower than that of a center‑fixed rod, yet faster than that of an end‑fixed rod, indicating that sliding partially releases the constraint imposed by the ring. The longest rotational relaxation time of the sliding rod exhibits a weak dependence on the mass distribution.

In synthesized rotaxane molecules, the interaction potential between the rod and the ring has been measured\cite{doi:10.1021/jacs.9b03792}. The molecular ring has to periodically overcome an activation energy to slide at the structural-unit scale, and the corresponding sliding diffusion coefficient exhibits an Arrhenius-type temperature dependence. Additionally, if contact friction between the rod and the ring is considered, extra energy dissipation occurs, which effectively modifies the translational friction coefficient $\zeta_\parallel$. Whether these effects matter depends on the length scale of the sliding-rod system. Furthermore, if the ring is released, the entire system gains an additional translational degree of freedom, in which case the coupling between translation and sliding would require further analysis.

\begin{acknowledgments}
We acknowledge the support of the National Natural Science Foundation of China (NSFC) (Nos. 22403021, 12247174, and 12274098), Wenzhou Institute, University of Chinese Academy of Sciences (Nos. WIUCASQD2022004, and WIUCASQD2021041), and Oujiang Laboratory (No. OJQDSP2022018).
\end{acknowledgments}

\section*{Author Declarations}
The authors have no conflicts to disclose.

\section*{Data Availability}
The data that supports the findings of this study are available within the article. 

\appendix

\section{Eigenfunction Expansion} \label{app:A}

We perform an eigenfunction expansion for the solution of Eq.~\eqref{smleq}, truncated up to $g$ terms
\begin{align}
	\tilde{\psi} (\tilde{s},\theta,\phi,\tilde{t}) = \tilde{\psi}_{\mathrm{eq}}^{1/2}(\tilde{s})\sum_{n=0}^{g}\sum_{l=0}^\infty\sum_{m=-l}^l A_{nlm}(\tilde{t}) \varphi_n(\tilde{s})Y_l^{m} (\theta,\phi), \label{solution_expansion}
\end{align}
where $Y_l^{m} (\theta,\phi)$ are the spherical harmonics, and $\varphi_n(\tilde{s})$ are defined as the solutions to the eigenvalue problem
\begin{equation} \label{eigenvalue_problem}
	\left\{
	\begin{aligned}
		&\mathcal{L}\varphi_n = \lambda_n \varphi_n, &&\tilde{s}\in \left(-\frac{1}{2},\frac{1}{2}\right), \\
		&\frac{\partial \varphi_n}{\partial \tilde{s}}+\frac{2}{1+4\alpha}\varphi_n= 0, &&\tilde{s} = - \frac{1}{2}, \\
		&\frac{\partial \varphi_n}{\partial \tilde{s}}-\frac{2}{1+4\alpha}\varphi_n= 0, &&\tilde{s} = \frac{1}{2},
	\end{aligned}
	\right.
\end{equation}
with $\mathcal{L}=-\partial^2/\partial \tilde{s}^2+\alpha/(\alpha+\tilde{s}^2)^2$. There exists one zero eigenvalue, whose corresponding eigenfunction is
\begin{align}
	\varphi_0 = (4\pi)^{1/2}\tilde{\psi}_{\mathrm{eq}}^{1/2}. \label{eigenfunction_0}
\end{align}
Other eigenfunctions $\varphi_n$ ($n=1,2,3,\cdots,g$) are obtained numerically. 

A weight function $\tilde{\psi}_{\mathrm{eq}}^{1/2}(\tilde{s})$ (see Eq.~\eqref{equilibrium_dimensionless}) is incorporated into the expansion Eq.~\eqref{solution_expansion}. This allows us to define the eigenvalue problem (Eq.~\eqref{eigenvalue_problem}) with a Hermitian operator $\mathcal{L}$\cite{risken1989fokker}. According to the Sturm-Liouville theory, we have the orthogonality relations
\begin{subequations} \label{orthogonality}
\begin{gather}
	\int _{-1/2}^{1/2}d\tilde{s}\, \varphi_n\varphi_{n'} = \delta_{nn'}, \label{orthogonality_varphi} \\
	\int_0^{2\pi} d\phi\int_0^{\pi} d\theta\,\sin\theta Y_l^{m}(\theta,\phi)Y_{l'}^{m'}(\theta,\phi)=\delta_{ll'}\delta_{mm'}, \label{orthogonality_sh}
\end{gather}
\end{subequations}
where $\delta_{ij}$ is the Kronecker delta. 

Substituting Eq.~\eqref{solution_expansion} into Eq.~\eqref{smleq} and using the orthogonality relation (Eq.~\eqref{orthogonality}), we obtain an equation for the coefficients $A_{nlm}(\tilde{t})$, i.e.,
\begin{align}
	\frac{\partial A_{n'lm}(\tilde{t})}{\partial \tilde{t}} = -\sum_{n=0}^{g} \mathcal{M}_{n'n}^lA_{nlm}(\tilde{t}), \label{eq_A0}
\end{align}
with $\mathcal{M}_{n'n}^l = \mathcal{M}_{nn'}^l$ and 
\begin{align}
	\mathcal{M}_{n'n}^l = \frac{1}{\beta_\parallel}\lambda_n\delta_{nn'}+ \int _{-1/2}^{1/2}d\tilde{s}\,\varphi_n(\tilde{s})\varphi_{n'}(\tilde{s})\frac{l(l+1)}{1+\beta_\perp \tilde{s}^2}. \label{MM}
\end{align}
The solution of Eq.~\eqref{eq_A0} is
\begin{align}
	A_{nlm}(\tilde{t}) = \sum_{p=0}^{g}\sum_{q=0}^{g}A_{qlm}(0)a_q^{lp}a_n^{lp}e^{-\lambda_{lp}\tilde{t}}, \label{coefficients_A}
\end{align}
where $\lambda_{lp}$ and $a_n^{lp}$ are eigenvalues and eigenvectors of the matrix $\mathcal{M}_{n'n}^l$, i.e.,
\begin{equation}
	\sum_{n=0}^{g} \mathcal{M}_{n'n}^la_n^{lp} = \lambda_{lp} a_{n'}^{lp},
\end{equation}
with the orthogonality relations
\begin{align}
	\sum_{n=0}^{g} a_n^{lp}a_n^{lq} = \delta_{pq}, \qquad \sum_{p=0}^{g} a_n^{lp}a_{n'}^{lp} = \delta_{nn'}.
\end{align}
In particular, for $l=0$, we have
\begin{equation}
	a_n^{0p} = \delta_{np} \label{eigenvector_0}.
\end{equation}
The coefficients $A_{qlm}(0)$ are determined from the initial distribution in Eq.~\eqref{initial} and used to construct the Green’s function given in Eq.~\eqref{Green'sFunction}.

\nocite{*}
\bibliography{aipsamp}

\end{document}